\documentclass[fleqn,twoside,twocolumn,nofootinbib,showkeys]{revtex4} % Specifies the document class %,unsortedaddress
\usepackage[english]{babel}
\usepackage[xdvi]{graphicx}
\usepackage{amssymb,amsmath}
\begin{document}

\title{Interplay of superconductivity and magnetism in FeSe$_{1-x}$Te$_x$ compounds.\\
Pressure effects.}

\author{A.S.~Panfilov}
\affiliation{B. Verkin Institute for Low Temperature Physics and
Engineering, National Academy of Sciences of Ukraine}%
\address{47, Lenin Ave., Kharkov 61103, Ukraine}%

\author{V.A. Pashchenko}
\affiliation{B. Verkin Institute for Low Temperature Physics and
Engineering, National Academy of Sciences of Ukraine}%
\address{47, Lenin Ave., Kharkov 61103, Ukraine}%

\author{G.E.~Grechnev}%1 автор
\affiliation{B. Verkin Institute for Low Temperature Physics and
Engineering, National Academy of Sciences of Ukraine}
\address{47, Lenin Ave., Kharkov 61103, Ukraine}
\email{grechnev@ilt.kharkov.ua}

\author{V.A.~Desnenko}
\affiliation{B. Verkin Institute for Low Temperature Physics and
Engineering, National Academy of Sciences of Ukraine}%
\address{47, Lenin Ave., Kharkov 61103, Ukraine}%

\author{A.V.~Fedorchenko}
\affiliation{B. Verkin Institute for Low Temperature Physics and
Engineering, National Academy of Sciences of Ukraine}
\address{47, Lenin Ave., Kharkov 61103, Ukraine}

\author{A.G.~Grechnev}%
\affiliation{B. Verkin Institute for Low Temperature Physics and
Engineering, National Academy of Sciences of Ukraine}%
\address{47, Lenin Ave., Kharkov 61103, Ukraine}%

\author{A.N. Bludov}
\affiliation{B. Verkin Institute for Low Temperature Physics and
Engineering, National Academy of Sciences of Ukraine}%
\address{47, Lenin Ave., Kharkov 61103, Ukraine}%

\author{S.L. Gnatchenko}
\affiliation{B. Verkin Institute for Low Temperature Physics and
Engineering, National Academy of Sciences of Ukraine}%
\address{47, Lenin Ave., Kharkov 61103, Ukraine}%

\author{D.A. Chareev}
\affiliation {Institute of Experimental Mineralogy, Russian
Academy of Sciences, Chernogolovka, Moscow Region 142432,
Russia}

\author{E.S. Mitrofanova}
\affiliation{Department of Materials Science, Moscow State University,
Moscow 119991, Russia}

\author{A.N. Vasiliev}
\affiliation{Low Temperature Physics and Superconductivity Department, Moscow State University, 
Moscow 119991, Russia,\\
Theoretical Physics and Applied Mathematics Department, Ural Federal
University, Mira Str. 19, 620002 Ekaterinburg, Russia}


\begin{abstract}
The influence of uniform pressures $P$ up to 5 kbar on
the superconducting transition temperature $T_c$ was studied for the
FeSe$_{1-x}$Te$_x$ ($x=0$, 0.85, 0.88 and 0.9) system.
For the first time, we observed a change in sign of the pressure effect on $T_c$ when going
from FeSe to tellurium rich alloys.
This has allowed to specify the pressure derivative
d$T_c$/d$P$ for
%FeSe$_{1-x}$Te$_x$
the system as a function of composition.
The observed dependence was compared with results of the {\em ab~initio}
%first principles
calculations of electronic structure and magnetism of FeSe, FeTe and
FeSe$_{0.5}$Te$_{0.5}$, and also with our recent experimental data
on pressure effects on magnetic susceptibilities of FeSe and FeTe
compounds in the normal state.
This comparison demonstrates a competing interplay between
superconductivity and magnetism in tellurium rich
FeSe$_{1-x}$Te$_x$ compounds.

\end{abstract}


\maketitle

\section{Introduction}
For the most families of recently discovered class of the Fe-based high-temperature
superconductors (HTSC) the emergence of superconductivity with doping or under uniform pressure 
is accompanied by suppression of the magnetic ordering
\cite{Lumsden,Chu,Paglione,Wen}.
It is widely believed then that spin fluctuations play an important role in formation of the Cooper pairs \cite{Mazin,Hirschfeld,Kohama}.
Nevertheless, as shown e.g. in Ref. \cite {Sadovskii}, for many Fe-based HTSCs the
experimental values of superconducting transition temperatures are well described
in the framework of the electron-phonon mechanism of pairing.
The close interrelation of magnetism and superconductivity determines the importance
of further studying of magnetic and superconducting properties and their evolution
under variations of composition, pressure, etc. for understanding
HTSC mechanism in the considered new class of iron compounds.
One of representatives of this class is the system of FeSe$_{1-x}$Te$_x$ chalcogenides,
which possesses the simplest crystal structure among iron-based superconductors,
that favors to experimental and theoretical studying the effects of chemical substitution
and high pressures on its properties.

Superconducting properties of FeSe$_{1-x}$Te$_x$ are characterized by nonmonotonic dependence
of transition temperatures $T_c$ on composition.
There is a noticeable growth from $T_c\simeq 8$~K for $x=0$ to the maximum value
$\sim 15$ K at $x\simeq 0.5$ with the subsequent falling to 0 K near $x\sim0.9$
(see, for example, Ref. \cite {Mizuguchi_2010a} and references therein).
Also, in FeSe compound the  extremely large rise of $T_c$ up to $35\div 37$ K
takes place with pressure $P\sim $70--80 kbar \cite{Braithwaite_2009,Medvedev}.
The similar behavior of $T_c$ under pressure was also observed in FeSe$_{0.5}$Te$_{0.5}$
compound \cite{Horigane_2009,Pietosa}.
With further increase of $x$ in FeSe$_{1-x}$Te$_x$ a tendency to reduction of the positive pressure effect is expected with even probable change of its sign, as it was observed in the related tellurium rich FeS$_{0.2}$Te$_{0.8}$ alloy \cite{Mizuguchi_2010a}.
This alleged change in sign of the pressure effect on $T_c$ in FeSe$_{1-x}$Te$_x$ under substitution of Te for Se could also explain the reason of unsuccessful attempts to observe superconductivity in FeTe under pressures up to 190 kbar \cite{Okada,Takahashi}.

Magnetic properties of FeSe$_{1-x}$Te$_x$ system were investigated in a number of works
\cite{Sales,Chen,Viennois,Yang,Noji,Liu,Fedorchenko1,Grechnev1,Ovchenkov}, however, data on
the magnetic susceptibility in the normal state remain incomplete and quantitatively inconsistent.
This is caused not only by a different quality of the samples used, but also by the existence
in them of impurities of iron and its secondary magnetic phases which considerably mask their
intrinsic magnetic susceptibility and must be carefully taken into account \cite{Fedorchenko1}.
The most adequate experimental data indicate that the susceptibility of FeSe$_{1-x}$Te$_x$
compounds increases gradually with Te content, being in FeTe about one order of magnitude lager then that of FeSe. Moreover, FeTe compound becomes magnetically unstable, and the antiferromagnetic ordering has been observed at temperatures about 70 K (see e.g. Ref.~\cite{Chen}).

It should be noted that the largest rise of magnetic susceptibility in the normal state,
$\chi(x)$, with increase of $x$ is observed in tellurium rich compounds, where, in turn,
the $T_c(x)$ dependence falls steeply down and FeTe compound is not superconductor under ambient conditions. This allows to assume a competing interplay between magnetism and superconductivity,
at least for this range of compositions.
In order to shed more light on the relationship between magnetic and superconducting properties in FeSe$_{1-x}$Te$_x$ system, it is very important to study evolution of these properties under high pressure. For this purpose in the present work we investigated the influence of hydrostatic pressure on the superconducting transition temperature, mainly in tellurium rich FeTe(Se) compounds. The obtained experimental results were compared with available data on behavior of magnetic susceptibility under pressure for the basic compounds FeSe \cite{Grechnev2} and FeTe
\cite{Fedorchenko2,Okada}, also supplemented by calculated pressure dependencies of
electronic structure and magnetic susceptibility for FeSe$_{0.5}$Te$_{0.5}$ compound.

\section{Experimental details and results}

The single crystals of FeSe$_{0.96}$ superconductor (hereinafter referred to as FeSe)
were grown during 50 days in evacuated quartz ampoules using the AlCl$_3$/KCl flux technique
with a constant temperature gradient along the ampoule length \cite{Chareev}.
Temperature of the hot end of the ampoule was kept at 427$^{\circ}$C,
when its more cold end was at about 380$^{\circ}$C.
A similar method was employed for the synthesis of tellurium-rich single crystals of
Fe$_{1+\delta}$Se$_{1-x}$Te$_x$ superconductors ($\delta\sim 0.05$, $x =$ 0.85, 0.88 and 0.90).
In this case we used the KCl/NaCl salt mixture and temperatures of the hot and cold ends
of the ampoule were 750$^{\circ}$C and about 700$^{\circ}$C, respectively.
The duration of the synthesis was 20-25 days.
Typical dimensions of the produced plate-like single crystals were
$(1-3)\times(1-3)\times(0.2-0.3)$ mm$^3$. Their tetragonal $P4/nmm$ structure was demonstrated at room temperature by an x-ray diffraction technique.
The crystals composition was determined using energy dispersive X-ray spectroscopy,
performed on a CAMECA SX100 (15 keV) analytical scanning electron microscope,
with an accuracy of the components ratio not worse than 2\%
(for details, see \cite{Chareev,Ovchenkov}).

The measurements of magnetic properties were performed using a SQUID magnetometer
(MPMS-XL5 Quantum Design) equipped with a miniature high-pressure cell of
a piston-cylinder type (similar to that described in Ref. \cite{Baran}).
The cell was made of non-magnetic CuBe alloy with the inside and outside diameters
of 1.6 mm and 5 mm, respectively.
Polyethylsiloxane liquid PES-3 was used as a hydrostatic pressure-transmitting medium.
The value of pressure at low temperatures was determined according to the known pressure
dependence of the superconducting transition temperature for a sample of pure tin
\cite{Jennings}, located inside the cell close to the measured sample.
The corresponding error did not exceed 0.2 kbar.

\begin{figure}[t]
\begin{center}
\includegraphics*[trim=0mm 0mm 0mm 0mm,scale=0.9]{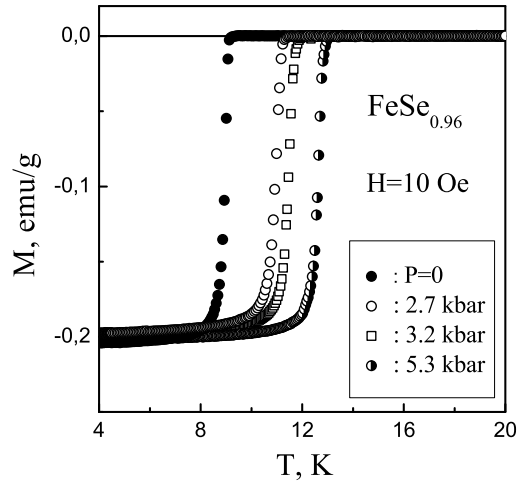}
\end{center}
\caption{\label{M(T)_FeSe}Temperature dependence of the magnetic moment of FeSe,
measured in magnetic field $H=10$ Oe at different pressures.}
\end{figure}

\begin{figure}[t]
\begin{center}
\includegraphics*[trim=0mm 0mm 0mm 0mm,scale=0.9]{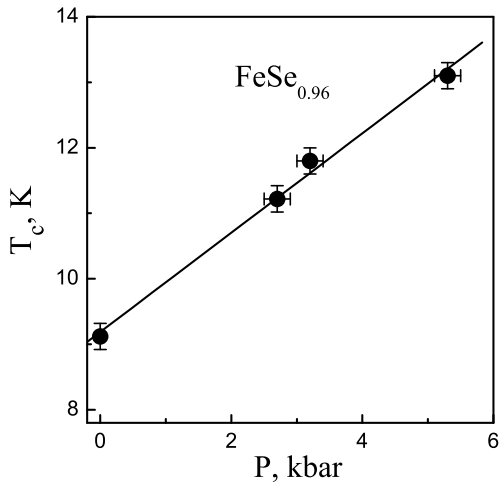}
\end{center}
\caption{\label{Tc(P)}Pressure dependence of the superconducting transition temperature for FeSe.}
\end{figure}

Fig. \ref{M(T)_FeSe} shows the temperature dependencies of magnetic moment $M(T)$ for FeSe
at different values of pressure, which were measured under cooling of the sample in zero
magnetic field (ZFC) followed by its heating in the field $H=10$ Oe.
Resulted from Fig. \ref{M(T)_FeSe} pressure dependence of the superconducting
transition temperature $T_c$, determined from here on by the onset of the transition,
is given in Fig. \ref{Tc(P)}.
Within the experimental errors and the operating range of pressure,
this dependence appeared to be close to linear that allows to evaluate the
pressure derivative d$T_c$/d$P$.

The $M(T)$ dependencies for tellurium-rich FeSe$_{1-x}$Te$_x$ compounds were measured at different pressures in ZFC regime, and are shown in Fig. \ref{M(T,P,x)}. They demonstrate clearly defined negative pressure effect on the superconducting transition temperature.
\begin{figure}[]
\begin{center}
\includegraphics*[trim=0mm 0mm 0mm 0mm,scale=0.9]{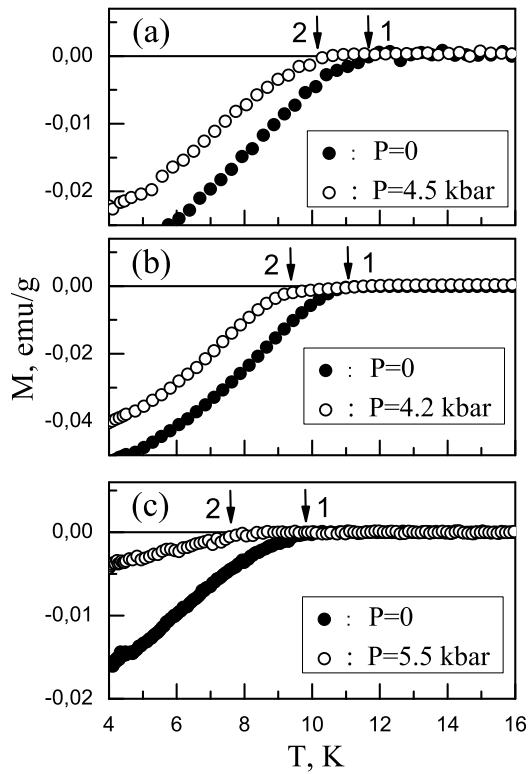}
\end{center}
\caption{\label{M(T,P,x)}Temperature dependencies of the magnetic moment measured in $H=10$ Oe
at two values of pressure for tellurium-rich FeSe$_{1-x}$Te$_x$ compounds:
(a) - $x = 0.85$, (b) - $x - 0,88$, (c) - $x =0.9$.
Arrows 1 and 2 denote $T_c$ at zero and finite values of pressure, respectively.}
\end{figure}
Experimental values of $T_c$ and its pressure derivative for all investigated samples
are listed in Table \ref{Experimental}.
As is evident from the presented data, the pressure effects on $T_c$ in the tellurium rich
FeSe$_{1-x}$Te$_x$ compounds are comparable in magnitude with that for FeSe but have
opposite negative sign.
\begin{table}[]
\caption{\label{Experimental} Superconducting transition temperature $T_c$ and its pressure derivative
d$T_c$/d$P$ for FeSe$_{1-x}$Te$_x$ compounds. }
\begin{center}
\begin{tabular}{ccc}
\hline\hline Composition & ~$T_c$ (K)~ &~~d$T_c$/d$P$ (K/kbar)~ \\
\hline
x= 0 & 9.12  & $0.78\pm 0.05$\\
x= 0.85 & 11.62 & $-0.31\pm 0.05$~~~\\
x= 0.88 & 11.05 & $-0.40\pm 0.05$~~~\\
x= 0.90 & 9.71  & $-0.40\pm 0.1$~~~\\
\hline\hline
\end{tabular}
\end{center}
\end{table}

\section{Calculations of electronic structure and magnetic susceptibility of
FeSe$_{0.5}$Te$_{0.5}$ compound}

\begin{figure}[]
\begin{center}
\includegraphics[scale=0.9]{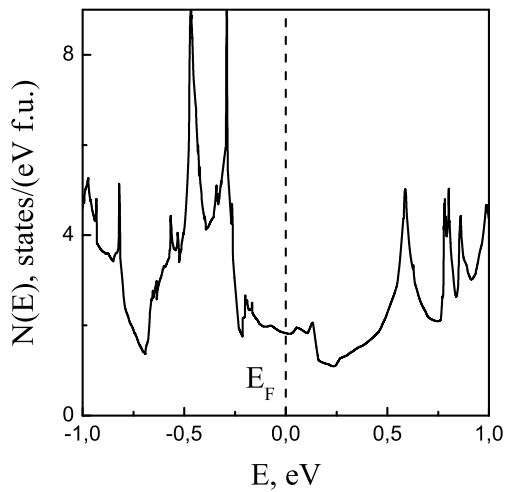}
\caption{\label{dos}Density of electronic states $N(E)$ of FeSe$_{0.5}$Te$_{0.5}$ compound.
The Fermi level position at 0 eV is marked by a vertical line.}
\end{center}
\end{figure}

For calculations of electronic structure of FeSe$_{0.5}$Te$_{0.5}$ compound we employed
the relativistic full potential LMTO method (FP-LMTO, RSPt implementation
\cite{wills10,Grechnev3}).
The exchange-correlation potential was treated within the local density approximation (LDA
\cite{barth72}) of the density functional theory (DFT).
The calculations were carried out for a supercell $2\times 2\times 1$, constructed by double
translations of the unit cell for the ordered tetragonal phase of FeSe and FeTe along
the crystallographic [100] and [010] directions, by using experimental values of crystal
lattice parameters for FeSe$_{0.5}$Te$_{0.5}$ from Refs. \cite {tsoi09,tsurkan11,malavi13}).
The calculated density of electronic states (DOS) $N(E)$ of the paramagnetic
FeSe$_{0.5}$Te$_{0.5}$ compound is presented in Fig.~\ref{dos}.
The Fermi level $E_{\rm F}$ is situated in the region of a local flat plateau of $N(E)$,
where the main contribution to DOS comes from the $d$-states of iron.
Such position of $E_{\rm F}$ implies a weak temperature dependence of the spin susceptibility
in FeSe$_{0.5}$Te$_{0.5}$, which is consistent with available experimental data
for this compound  \cite{Sales,Fedorchenko1,Grechnev1}.

To evaluate the paramagnetic susceptibility of FeSe$_{0.5}$Te$_{0.5}$ compound,
the FP-LMTO calculations of field-induced spin and orbital (Van Vleck) magnetic moments were carried out with the approach described in Ref. \cite{Grechnev3} within the local
spin density approximation (LSDA) of DFT.
The relativistic effects, including spin-orbit coupling, were incorporated,
and the effect of an external magnetic field $\mathbf{B}$ was taken into account
self-consistently by means of the Zeeman term:
\begin{equation}
\label{zeeman}
{\cal H}_{Z}=\mu_{B} \mathbf{B}\cdot (2\hat{\bf{s}}+\hat{\bf{l}}) ,
\end{equation}
Here $\mu_{\rm B}$ is the Bohr magneton, $\hat{\bf{s}}$ and $\hat{\bf{l}}$
are the spin and orbital angular momentum operators, respectively.
The ratio of the field-induced magnetizations to the field strength
($\mathbf{B}=10$~T) provided corresponding spin and orbital components of magnetic
susceptibilities, $\chi_{\rm spin}$ and $\chi_{\rm orb}$, respectively.

According to results of the calculations, the exchange-enhanced spin paramagnetism
$\chi_{\rm spin}$ appears to be the main contribution to magnetic susceptibility
of FeSe$_{0.5}$Te$_{0.5}$ compound.
Within the Stoner model, this contribution can be presented as:
$\chi_{\rm spin}=S \mu^2_{\rm B}N(E_{\rm F})$, where $S$ is the Stoner factor, $N(E_{\rm F})$ DOS at the Fermi level, $\mu_{\rm B}$ the Bohr magneton.
Using the calculated values of spin magnetic susceptibility of FeSe$_{0.5}$Te$_{0.5}$ compound,
$\chi_{\rm spin} \simeq 0.6\times 10^{-3}$ emu/mol, and DOS at the Fermi level,
$N(E_{\rm F})\simeq 1.85$ eV$^{-1}$, we have obtained the estimation of the Stoner factor:
$S\simeq $10.
It should be noted that the above listed calculated value of $\chi_{\rm spin}$
is in agreement with the experimental magnetic susceptibility of FeSe$_{0.5}$Te$_{0.5}$
compound in the normal state (see Refs. \cite{Sales,Grechnev1}).
This confirms the dominating role of the spin contribution to magnetism of
FeSe$_{0.5}$Te$_{0.5}$ compound, that is, apparently, characteristic for
the whole FeSe$_{1-x}$Te$_{x}$ system \cite{Grechnev1,Grechnev2,Fedorchenko2}.

By using the experimental data of Ref. \cite{malavi13} on evaluation of the lattice parameters
of FeSe$_{0.5}$Te$_{0.5}$ under uniform compression, we calculated the behavior of
density of electronic states at the Fermi level.
For the region of small pressures (0$\div$10 kbar) we established the growth of $N(E_{\rm F})$
with the rate of d$\ln N(E_{\rm F})/{\rm d}P\simeq 1 $ Mbar$^{-1}$.
We should note that such behavior of $N(E_{\rm F})$ correlates with increase
of the superconducting transition temperature in FeSe$_{0.5}$Te$_{0.5}$
under pressure \cite{Horigane_2009,Pietosa}.

Within the considered above method of calculation of magnetic susceptibility,
we also investigated the dependence of $\chi$ in FeSe$_{0.5}$Te$_{0.5}$ compound on the uniform pressure.
By direct calculations of the field-induced magnetic moments, we have obtained
the value of pressure derivative of paramagnetic susceptibility,
d\,ln$\chi$/d$P\simeq 13$ Mbar$^{-1}$, which appeared to be
close to the corresponding values in FeSe and FeTe (see Tab. \ref{dlnX/dP}).
In order to clarify the mechanism of the strong increase of magnetic susceptibility in
FeSe$_ {0.5}$Te$_{0.5}$ under pressure, we calculated value of $\chi$ as a function of the unit cell volume $V$ and the internal structural parameter $Z$,
which determines the relative height of chalcogen atoms over the plane of iron atoms.
Then the corresponding pressure effect on $\chi$ can be presented as follows:
\begin{eqnarray}
{{\rm d\,ln}\chi\over {\rm d}P}= {\partial\,{\rm ln}\chi\over
\partial\,{\rm ln}V}\times{{\rm d\,ln}V\over {\rm d}P}+{\partial\,{\rm ln}\chi\over
\partial Z}\times{{\rm d}Z\over {\rm d}P}.
\label{dX/dP}
\end{eqnarray}
By small variations of the cell volume $V$ and the structural parameter $Z$
near their experimental values, the following partial derivatives
of paramagnetic susceptibility for FeSe$_{0.5}$Te$_{0.5}$ were calculated to be
$\partial\,{\rm ln}\chi/\partial\,{\rm ln}V\simeq 10$ and
$\partial\,{\rm ln}\chi/\partial Z\simeq 90$.
The necessary values for the compressibility of FeSe$_{0.5}$Te$_{0.5}$,
 d\,ln$V$/d$P=-3.1$ Mbar$^{-1}$, and behavior of parameter $Z$ under pressure,
d$Z$/d$P\simeq 0.49$ Mbar$^{-1}$, were taken from Ref. \cite{malavi13}.
By substituting the values of these parameters in Eq. (\ref{dX/dP}) we have found that the calculated in this work large positive pressure effect on $\chi$ in FeSe$_{0.5}$Te$_{0.5}$ is related to the strong sensitivity of susceptibility to the parameter $Z$ and its change under pressure, that determines the dominant positive contribution.

\section{Discussion}
\begin{figure}[t]
\begin{center}
\includegraphics*[trim=0mm 0mm 0mm 0mm,scale=0.9]{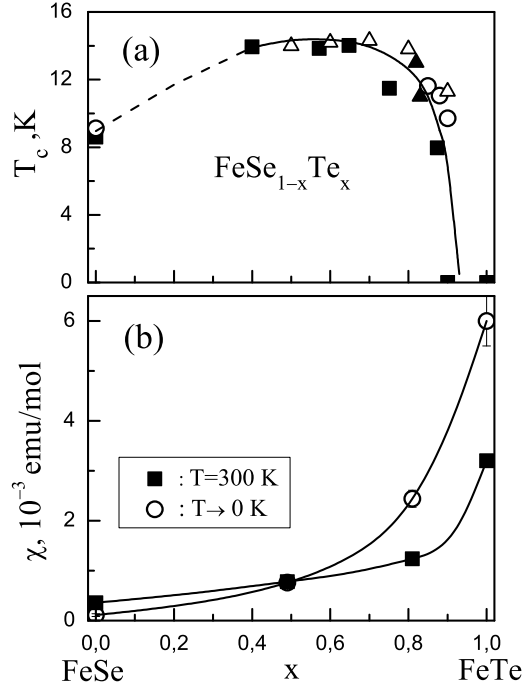}
\end{center}
\caption{\label{phase'diagramm}Concentration dependencies of (a) superconducting transition temperature (from Refs:~$\blacksquare~ -$ \cite{Mizuguchi_2010a}, $\triangle~ -$
\cite{Noji}, {\large$\blacktriangle$} $-$ \cite{Koshika},
{\Large$\circ$} $-$ present work) and (b) magnetic susceptibility in
the normal state for $T\to 0$ K and $T=300$ K (from Ref.
\cite{Grechnev1}) on Te content $x$ in FeSe$_{1-x}$Te$_x$ compounds.}
\end{figure}

Experimental values of superconducting transition temperatures for the investigated in
this work compounds are in agreement with the literature data (see Fig.~\ref{phase'diagramm}a).
The most studied range of compositions ($x\ge 0.4$) is characterized by the sharp reduction of $T_c$ with increasing $x$ at $x\ge 0.7$, and the total disappearance of superconductivity for $x\to 1$. In the same range of compositions the strong growth of magnetic susceptibility in the normal
state was observed (Fig. \ref{phase'diagramm}b).
The obtained strictly opposite tendencies in composition dependencies of superconductivity
and magnetism in FeSe$_ {1-x}$Te$_x$ system allow to assume that interrelation of these
phenomena has competing character, at least for the tellurium rich compounds.

\begin{figure}[]
\begin{center}
\includegraphics*[scale=0.9]{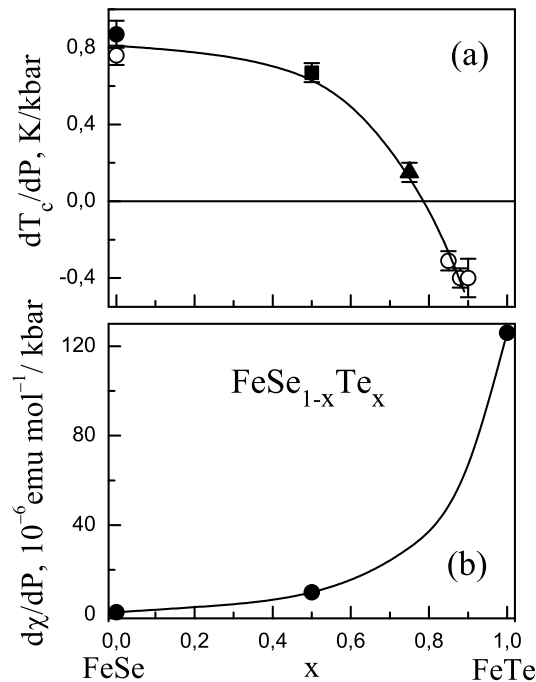}
\end{center}
\caption{\label{dTc/dP} (a): The values of d$T_c$/d$P$ derivative
depending on Te composition $x$ in FeSe$_{1-x}$Te$_x$ compounds
({\Large$\circ$} $-$ this work; {\Large$\bullet$} $-$
\cite{Bendele_2012}; $\blacksquare~ -$ \cite{Pietosa};
{\large$\blacktriangle$} $-$ \cite{Mizuguchi_2010b}.
(b): dependence of the pressure derivative of magnetic susceptibility in the
normal state on Te composition $x$ (see more details in the text).}
\end{figure}

Let us consider now the evolution of superconducting and magnetic properties of FeSe$_{1-x}$Te$_x$ compounds under uniform pressure. Experimental values for pressure derivatives of the
superconducting transition temperature are given in Fig.~\ref{dTc/dP}a,
which include the known published data and the results of this work.
Apparently, the available data describe the monotonous reduction of the pressure effect in $T_c$
in process of selenium substitution with tellurium, and the change of its sign at $x\sim 0.8$.
This trend is also consistent with the value of  d$T_c$/d$P\simeq -0.25$ K/kbar for
the related FeS$_{0.2}$Te$_{0.8}$ compound \cite{Mizuguchi_2010a}.

Unlike the pressure effect on $T_c$, which changes its sign as a function of composition
(Fig.~\ref{dTc/dP}a), the magnetic susceptibility of FeSe(Te) system in the normal state
is characterized by substantial growth under pressure for the whole system.
This conclusion follows from available experimental data and theoretical estimates for the
basic FeSe \cite{Grechnev2} and FeTe \cite{Fedorchenko2,Okada} compounds, together with the results of present calculations for pressure dependence of magnetic susceptibility in
FeSe$_{0.5}$Te$_{0.5}$ compound.

As can be seen from the values of pressure derivative of susceptibility,
d\,ln$\chi$/d$P$, given in Table \ref{dlnX/dP},
for considered FeSe(Te) system the pressure effect not only much exceeds its
typical value in the exchange-enhanced itinerant paramagnets \cite {Grechnev3},
but also has the opposite {\em positive} sign.
This implies an unusual for metallic system possibility of transition to the ferromagnetic
state under the influence of experimentally achievable pressures.
This is particularly the case of FeTe compound where the pressure effect is the largest.
In Ref.~\cite{Fedorchenko2} from the analysis of temperature dependence of susceptibility for
FeTe in the paramagnetic region within the Curie-Weiss law,
the values of the paramagnetic Curie temperature and its pressure derivative were evaluated to be $\Theta\simeq -240$ K and d$\Theta$/d$P\sim 7$ K/kbar.
Corresponding to them rough estimate of the critical pressure for ferromagnetic transition
amounts to 35 kbar.
This is in reasonable agreement with results of Ref. \cite{Bendele2013},
where the ferromagnetic state was observed in FeTe for the first time
under pressures of $P\ge 20$ kbar.

\begin{table}[]
\caption{\label{dlnX/dP} Pressure derivatives of the magnetic susceptibility,
d\,ln$\chi$/d$P$, for FeSe$_{1-x}$Te$_x$  compounds.
Experimental temperatures are specified in brackets,
results of calculations correspond to $T=0$ K.
The data for FeTe are referred to the paramagnetic state.}
\vspace{3pt}
\begin{center}
\begin{tabular}{lcc}
\hline\hline
Compound &  \multicolumn{2}{c}{d\,ln$\chi$/d$P$, Mbar$^{-1}$} \\
\hline &   Experiment & Theory \\
 \vspace{-0pt}
 FeSe                   &  $10\pm 3$ (78 K)$^{\rm a}$  & $\simeq 8^{\rm a}$ \\
                        &  $\sim 9$ (20 K)$~^{\rm b}$ &            \\
 FeSe$_{0.5}$Te$_{0.5}$ & $-$ & $\simeq 13$ \\
 FeTe                   & ~~   $23 \pm 1.5$ (78 K)$~^{\rm c}$~~  & $\sim 20~^{\rm c}$ \\
                        &  $\simeq21$ (78 K)~$ ^{\rm d}$     &           \\
\hline\hline
\end{tabular}
\end{center}
%\vspace{3pt}
$^{\rm a}-$ from Ref. \cite{Grechnev2}, $^{\rm b}-$ from NMR data %(Knight shift)
 of Ref. \cite{Imai},
$^{\rm c}-$ from Ref. \cite{Fedorchenko2}, $^{\rm d}-$ from magnetization data of Ref.
\cite{Okada}.
\end{table}

For convenient comparison of the observed pressure effects in superconducting transition
temperatures (Fig.~\ref{dTc/dP}a) with pressure effects in magnetic susceptibility,
the values of pressure derivatives of susceptibility,
d$\chi$/d$P\equiv\chi\times$d\,ln$\chi$/d$P$, are presented in Fig.~\ref{dTc/dP}b
for FeSe, FeSe$_{0.5}$Te$_{0.5}$ and FeTe.
To evaluate these derivatives we used the corresponding values of $\chi(T\to 0$ K)
from Ref.~\cite{Grechnev1} (Fig.~\ref{phase'diagramm}b) and the average values
of d\,ln$\chi$/d$P$ from Table~\ref{dlnX/dP}.
As can be seen in Fig.~\ref{dTc/dP}, the presented composition dependencies of pressure effects
in magnetic and superconducting properties of FeSe$_{1-x}$Te$_x$ system are
strictly opposite to one another.
This fact, along with similar trends in behavior of magnetic susceptibility and $T_c$
as function of composition at ambient pressure (Fig.~\ref{phase'diagramm}), specifies on
antagonistic interrelation of magnetism and superconductivity in  FeSe$_{1-x}$Te$_x$ system,
which is most pronounced in the tellurium rich compounds.

\section*{Conclusions}
In this work the negative pressure effect on the superconducting transition temperature of
tellurium rich FeSe$_{1-x}$Te$_x$ compounds was observed for the first time.
The obtained data allowed to establish an overall picture of the composition dependence for
the pressure effect on $T_c$, which monotonously decreases with growth of $x$
and changes its sign at $x\sim 0.8$.

Another feature of FeSe$_{1-x}$Te$_x$ compounds is anomalously large and positive
pressure effect on magnetic susceptibility in the normal state for all compositions,
which grows with substitution of tellurium for selenium.
As appears from the present calculations of the pressure effect on $\chi$ for
FeSe$_{0.5}$Te$_{0.5}$ and the earlier similar calculations for FeSe and FeTe,
the large positive pressure effect on susceptibility in FeSe$_ {1-x}$Te$_x$ compounds
is determined by the dominating positive contribution caused by the strong
sensitivity of paramagnetic susceptibility to internal structural parameter $Z$
and its change under pressure.
It should be noted that the largest pressure effect on $\chi$ appears in FeTe compound, and
that is a source of the observed its ferromagnetic state at high pressures \cite{Bendele2013}.

Finally, the revealed here opposite trends in composition and pressure dependencies
of superconducting transition temperature and magnetic susceptibility in the normal state
indicate to antagonistic interrelation between superconductivity and magnetism in
FeSe$_{1-x}$Te$_x$ chalcogenides.
This tendency obviously has to be taken into account in further studies
of possible role of magnetic excitations in the mechanism of superconductivity
in Fe-based HTSCs.

\begin{acknowledgments}

Authors express deep gratitude to Professor G.G. Levchenko for
providing us with the high-pressure cell for measurements and
his kind advices on its operation and maintenance.

This work was supported by the Russian-Ukrainian RFBR-NASU
project 01-02-12.

\end{acknowledgments}

\end{document}